\documentclass[aps,prd,twocolumn,showpacs]{revtex4}
\usepackage{graphicx,color}
\usepackage{latexsym}
\usepackage{amsmath}
\usepackage{amsfonts}
\usepackage{amssymb}
\usepackage[latin1]{inputenc}

\def\m{\mu}

\def\r{\rho}

\def\t{\tau}

\def\x{\xi}

\def\L{\Lambda}

\newcommand{\be}{\begin{equation}}
\newcommand{\ee}{\end{equation}}
\newcommand{\bea}{\begin{eqnarray}}
\newcommand{\eea}{\end{eqnarray}}

\newcommand{\pa}{\partial}

\newcommand{\bb}{\bibitem}
\def\pls{\partial\!\!\!/}
\def\bb{\bibitem}
\def\as{A\!\!\!/}
\def\cas{{\cal A}\!\!\!/}
\def\cvs{{\cal V}\!\!\!/}
\def\ps{p\!\!\!/}
\def\bs{b\!\!\!/}

\def\n{\nu}
\def\m{\mu}
\def\n{\nu}
\def\bb{\bibitem}

\newcommand{\ben}{\begin{eqnarray}}
\newcommand{\een}{\end{eqnarray}}

\begin{document}
\immediate\write16{<WARNING: FEYNMAN macros work only with emTeX-dvivers
                    (dviscr.exe, dvihplj.exe, dvidot.exe, etc.) >}
\newdimen\Lengthunit
\newcount\Nhalfperiods
\Lengthunit = 1.5cm
\Nhalfperiods = 9
\catcode`\*=11
\newdimen\L*   \newdimen\d*   \newdimen\d**
\newdimen\dm*  \newdimen\dd*  \newdimen\dt*
\newdimen\a*   \newdimen\b*   \newdimen\c*
\newdimen\a**  \newdimen\b**
\newdimen\xL*  \newdimen\yL*
\newcount\k*   \newcount\l*   \newcount\m*
\newcount\n*   \newcount\dn*  \newcount\r*
\newcount\N*   \newcount\*one \newcount\*two  \*one=1 \*two=2
\newcount\*ths \*ths=1000
\def\GRAPH(hsize=#1)#2{\hbox to #1\Lengthunit{#2\hss}}
\def\Linewidth#1{\special{em:linewidth #1}}
\Linewidth{.4pt}
\def\sm*{\special{em:moveto}}
\def\sl*{\special{em:lineto}}
\newbox\spm*   \newbox\spl*
\setbox\spm*\hbox{\sm*}
\setbox\spl*\hbox{\sl*}
\def\mov#1(#2,#3)#4{\rlap{\L*=#1\Lengthunit\kern#2\L*\raise#3\L*\hbox{#4}}}
\def\smov#1(#2,#3)#4{\rlap{\L*=#1\Lengthunit
\xL*=\xscale\L*\yL*=\yscale\L*\kern#2\xL*\raise#3\yL*\hbox{#4}}}
\def\mov*(#1,#2)#3{\rlap{\kern#1\raise#2\hbox{#3}}}
\def\lin#1(#2,#3){\rlap{\sm*\mov#1(#2,#3){\sl*}}}
\def\arr*(#1,#2,#3){\mov*(#1\dd*,#1\dt*){%
\sm*\mov*(#2\dd*,#2\dt*){\mov*(#3\dt*,-#3\dd*){\sl*}}%
\sm*\mov*(#2\dd*,#2\dt*){\mov*(-#3\dt*,#3\dd*){\sl*}}}}
\def\arrow#1(#2,#3){\rlap{\lin#1(#2,#3)\mov#1(#2,#3){%
\d**=-.012\Lengthunit\dd*=#2\d**\dt*=#3\d**%
\arr*(1,10,4)\arr*(3,8,4)\arr*(4.8,4.2,3)}}}
\def\arrlin#1(#2,#3){\rlap{\L*=#1\Lengthunit\L*=.5\L*%
\lin#1(#2,#3)\mov*(#2\L*,#3\L*){\arrow.1(#2,#3)}}}
\def\dasharrow#1(#2,#3){\rlap{%
{\Lengthunit=0.9\Lengthunit\dashlin#1(#2,#3)\mov#1(#2,#3){\sm*}}%
\mov#1(#2,#3){\sl*\d**=-.012\Lengthunit\dd*=#2\d**\dt*=#3\d**%
\arr*(1,10,4)\arr*(3,8,4)\arr*(4.8,4.2,3)}}}
\def\clap#1{\hbox to 0pt{\hss #1\hss}}
\def\ind(#1,#2)#3{\rlap{%
\d*=.1\Lengthunit\kern#1\d*\raise#2\d*\hbox{\lower2pt\clap{$#3$}}}}
\def\sh*(#1,#2)#3{\rlap{%
\dm*=\the\n*\d**\xL*=\xscale\dm*\yL*=\yscale\dm*
\kern#1\xL*\raise#2\yL*\hbox{#3}}}
\def\calcnum*#1(#2,#3){\a*=1000sp\b*=1000sp\a*=#2\a*\b*=#3\b*%
\ifdim\a*<0pt\a*-\a*\fi\ifdim\b*<0pt\b*-\b*\fi%
\ifdim\a*>\b*\c*=.96\a*\advance\c*.4\b*%
\else\c*=.96\b*\advance\c*.4\a*\fi%
\k*\a*\multiply\k*\k*\l*\b*\multiply\l*\l*%
\m*\k*\advance\m*\l*\n*\c*\r*\n*\multiply\n*\n*%
\dn*\m*\advance\dn*-\n*\divide\dn*2\divide\dn*\r*%
\advance\r*\dn*%
\c*=\the\Nhalfperiods5sp\c*=#1\c*\ifdim\c*<0pt\c*-\c*\fi%
\multiply\c*\r*\N*\c*\divide\N*10000}
\def\dashlin#1(#2,#3){\rlap{\calcnum*#1(#2,#3)%
\d**=#1\Lengthunit\ifdim\d**<0pt\d**-\d**\fi%
\divide\N*2\multiply\N*2\advance\N*1%
\divide\d**\N*\sm*\n*\*one\sh*(#2,#3){\sl*}%
\loop\advance\n*\*one\sh*(#2,#3){\sm*}\advance\n*\*one\sh*(#2,#3){\sl*}%
\ifnum\n*<\N*\repeat}}
\def\dashdotlin#1(#2,#3){\rlap{\calcnum*#1(#2,#3)%
\d**=#1\Lengthunit\ifdim\d**<0pt\d**-\d**\fi%
\divide\N*2\multiply\N*2\advance\N*1\multiply\N*2%
\divide\d**\N*\sm*\n*\*two\sh*(#2,#3){\sl*}\loop%
\advance\n*\*one\sh*(#2,#3){\kern-1.48pt\lower.5pt\hbox{\rm.}}%
\advance\n*\*one\sh*(#2,#3){\sm*}%
\advance\n*\*two\sh*(#2,#3){\sl*}\ifnum\n*<\N*\repeat}}
\def\shl*(#1,#2)#3{\kern#1#3\lower#2#3\hbox{\unhcopy\spl*}}
\def\trianglin#1(#2,#3){\rlap{\toks0={#2}\toks1={#3}\calcnum*#1(#2,#3)%
\dd*=.57\Lengthunit\dd*=#1\dd*\divide\dd*\N*%
\d**=#1\Lengthunit\ifdim\d**<0pt\d**-\d**\fi%
\multiply\N*2\divide\d**\N*\advance\N*-1\sm*\n*\*one\loop%
\shl**{\dd*}\dd*-\dd*\advance\n*2%
\ifnum\n*<\N*\repeat\n*\N*\advance\n*1\shl**{0pt}}}
\def\wavelin#1(#2,#3){\rlap{\toks0={#2}\toks1={#3}\calcnum*#1(#2,#3)%
\dd*=.23\Lengthunit\dd*=#1\dd*\divide\dd*\N*%
\d**=#1\Lengthunit\ifdim\d**<0pt\d**-\d**\fi%
\multiply\N*4\divide\d**\N*\sm*\n*\*one\loop%
\shl**{\dd*}\dt*=1.3\dd*\advance\n*1%
\shl**{\dt*}\advance\n*\*one%
\shl**{\dd*}\advance\n*\*two%
\dd*-\dd*\ifnum\n*<\N*\repeat\n*\N*\shl**{0pt}}}
\def\w*lin(#1,#2){\rlap{\toks0={#1}\toks1={#2}\d**=\Lengthunit\dd*=-.12\d**%
\N*8\divide\d**\N*\sm*\n*\*one\loop%
\shl**{\dd*}\dt*=1.3\dd*\advance\n*\*one%
\shl**{\dt*}\advance\n*\*one%
\shl**{\dd*}\advance\n*\*one%
\shl**{0pt}\dd*-\dd*\advance\n*1\ifnum\n*<\N*\repeat}}
\def\l*arc(#1,#2)[#3][#4]{\rlap{\toks0={#1}\toks1={#2}\d**=\Lengthunit%
\dd*=#3.037\d**\dd*=#4\dd*\dt*=#3.049\d**\dt*=#4\dt*\ifdim\d**>16mm%
\d**=.25\d**\n*\*one\shl**{-\dd*}\n*\*two\shl**{-\dt*}\n*3\relax%
\shl**{-\dd*}\n*4\relax\shl**{0pt}\else\ifdim\d**>5mm%
\d**=.5\d**\n*\*one\shl**{-\dt*}\n*\*two\shl**{0pt}%
\else\n*\*one\shl**{0pt}\fi\fi}}
\def\d*arc(#1,#2)[#3][#4]{\rlap{\toks0={#1}\toks1={#2}\d**=\Lengthunit%
\dd*=#3.037\d**\dd*=#4\dd*\d**=.25\d**\sm*\n*\*one\shl**{-\dd*}%
\n*3\relax\sh*(#1,#2){\xL*=\xscale\dd*\yL*=\yscale\dd*
\kern#2\xL*\lower#1\yL*\hbox{\sm*}}%
\n*4\relax\shl**{0pt}}}
\def\arc#1[#2][#3]{\rlap{\Lengthunit=#1\Lengthunit%
\sm*\l*arc(#2.1914,#3.0381)[#2][#3]%
\smov(#2.1914,#3.0381){\l*arc(#2.1622,#3.1084)[#2][#3]}%
\smov(#2.3536,#3.1465){\l*arc(#2.1084,#3.1622)[#2][#3]}%
\smov(#2.4619,#3.3086){\l*arc(#2.0381,#3.1914)[#2][#3]}}}
\def\dasharc#1[#2][#3]{\rlap{\Lengthunit=#1\Lengthunit%
\d*arc(#2.1914,#3.0381)[#2][#3]%
\smov(#2.1914,#3.0381){\d*arc(#2.1622,#3.1084)[#2][#3]}%
\smov(#2.3536,#3.1465){\d*arc(#2.1084,#3.1622)[#2][#3]}%
\smov(#2.4619,#3.3086){\d*arc(#2.0381,#3.1914)[#2][#3]}}}
\def\wavearc#1[#2][#3]{\rlap{\Lengthunit=#1\Lengthunit%
\w*lin(#2.1914,#3.0381)%
\smov(#2.1914,#3.0381){\w*lin(#2.1622,#3.1084)}%
\smov(#2.3536,#3.1465){\w*lin(#2.1084,#3.1622)}%
\smov(#2.4619,#3.3086){\w*lin(#2.0381,#3.1914)}}}
\def\shl**#1{\c*=\the\n*\d**\d*=#1%
\a*=\the\toks0\c*\b*=\the\toks1\d*\advance\a*-\b*%
\b*=\the\toks1\c*\d*=\the\toks0\d*\advance\b*\d*%
\a*=\xscale\a*\b*=\yscale\b*%
\raise\b*\rlap{\kern\a*\unhcopy\spl*}}
\def\wlin*#1(#2,#3)[#4]{\rlap{\toks0={#2}\toks1={#3}%
\c*=#1\l*\c*\c*=.01\Lengthunit\m*\c*\divide\l*\m*%
\c*=\the\Nhalfperiods5sp\multiply\c*\l*\N*\c*\divide\N*\*ths%
\divide\N*2\multiply\N*2\advance\N*1%
\dd*=.002\Lengthunit\dd*=#4\dd*\multiply\dd*\l*\divide\dd*\N*%
\d**=#1\multiply\N*4\divide\d**\N*\sm*\n*\*one\loop%
\shl**{\dd*}\dt*=1.3\dd*\advance\n*\*one%
\shl**{\dt*}\advance\n*\*one%
\shl**{\dd*}\advance\n*\*two%
\dd*-\dd*\ifnum\n*<\N*\repeat\n*\N*\shl**{0pt}}}
\def\wavebox#1{\setbox0\hbox{#1}%
\a*=\wd0\advance\a*14pt\b*=\ht0\advance\b*\dp0\advance\b*14pt%
\hbox{\kern9pt%
\mov*(0pt,\ht0){\mov*(-7pt,7pt){\wlin*\a*(1,0)[+]\wlin*\b*(0,-1)[-]}}%
\mov*(\wd0,-\dp0){\mov*(7pt,-7pt){\wlin*\a*(-1,0)[+]\wlin*\b*(0,1)[-]}}%
\box0\kern9pt}}
\def\rectangle#1(#2,#3){%
\lin#1(#2,0)\lin#1(0,#3)\mov#1(0,#3){\lin#1(#2,0)}\mov#1(#2,0){\lin#1(0,#3)}}
\def\dashrectangle#1(#2,#3){\dashlin#1(#2,0)\dashlin#1(0,#3)%
\mov#1(0,#3){\dashlin#1(#2,0)}\mov#1(#2,0){\dashlin#1(0,#3)}}
\def\waverectangle#1(#2,#3){\L*=#1\Lengthunit\a*=#2\L*\b*=#3\L*%
\ifdim\a*<0pt\a*-\a*\def\x*{-1}\else\def\x*{1}\fi%
\ifdim\b*<0pt\b*-\b*\def\y*{-1}\else\def\y*{1}\fi%
\wlin*\a*(\x*,0)[-]\wlin*\b*(0,\y*)[+]%
\mov#1(0,#3){\wlin*\a*(\x*,0)[+]}\mov#1(#2,0){\wlin*\b*(0,\y*)[-]}}
\def\calcparab*{%
\ifnum\n*>\m*\k*\N*\advance\k*-\n*\else\k*\n*\fi%
\a*=\the\k* sp\a*=10\a*\b*\dm*\advance\b*-\a*\k*\b*%
\a*=\the\*ths\b*\divide\a*\l*\multiply\a*\k*%
\divide\a*\l*\k*\*ths\r*\a*\advance\k*-\r*%
\dt*=\the\k*\L*}
\def\arcto#1(#2,#3)[#4]{\rlap{\toks0={#2}\toks1={#3}\calcnum*#1(#2,#3)%
\dm*=135sp\dm*=#1\dm*\d**=#1\Lengthunit\ifdim\dm*<0pt\dm*-\dm*\fi%
\multiply\dm*\r*\a*=.3\dm*\a*=#4\a*\ifdim\a*<0pt\a*-\a*\fi%
\advance\dm*\a*\N*\dm*\divide\N*10000%
\divide\N*2\multiply\N*2\advance\N*1%
\L*=-.25\d**\L*=#4\L*\divide\d**\N*\divide\L*\*ths%
\m*\N*\divide\m*2\dm*=\the\m*5sp\l*\dm*%
\sm*\n*\*one\loop\calcparab*\shl**{-\dt*}%
\advance\n*1\ifnum\n*<\N*\repeat}}
\def\arrarcto#1(#2,#3)[#4]{\L*=#1\Lengthunit\L*=.54\L*%
\arcto#1(#2,#3)[#4]\mov*(#2\L*,#3\L*){\d*=.457\L*\d*=#4\d*\d**-\d*%
\mov*(#3\d**,#2\d*){\arrow.02(#2,#3)}}}
\def\dasharcto#1(#2,#3)[#4]{\rlap{\toks0={#2}\toks1={#3}\calcnum*#1(#2,#3)%
\dm*=\the\N*5sp\a*=.3\dm*\a*=#4\a*\ifdim\a*<0pt\a*-\a*\fi%
\advance\dm*\a*\N*\dm*%
\divide\N*20\multiply\N*2\advance\N*1\d**=#1\Lengthunit%
\L*=-.25\d**\L*=#4\L*\divide\d**\N*\divide\L*\*ths%
\m*\N*\divide\m*2\dm*=\the\m*5sp\l*\dm*%
\sm*\n*\*one\loop%
\calcparab*\shl**{-\dt*}\advance\n*1%
\ifnum\n*>\N*\else\calcparab*%
\sh*(#2,#3){\kern#3\dt*\lower#2\dt*\hbox{\sm*}}\fi%
\advance\n*1\ifnum\n*<\N*\repeat}}
\def\*shl*#1{%
\c*=\the\n*\d**\advance\c*#1\a**\d*\dt*\advance\d*#1\b**%
\a*=\the\toks0\c*\b*=\the\toks1\d*\advance\a*-\b*%
\b*=\the\toks1\c*\d*=\the\toks0\d*\advance\b*\d*%
\raise\b*\rlap{\kern\a*\unhcopy\spl*}}
\def\calcnormal*#1{%
\b**=10000sp\a**\b**\k*\n*\advance\k*-\m*%
\multiply\a**\k*\divide\a**\m*\a**=#1\a**\ifdim\a**<0pt\a**-\a**\fi%
\ifdim\a**>\b**\d*=.96\a**\advance\d*.4\b**%
\else\d*=.96\b**\advance\d*.4\a**\fi%
\d*=.01\d*\r*\d*\divide\a**\r*\divide\b**\r*%
\ifnum\k*<0\a**-\a**\fi\d*=#1\d*\ifdim\d*<0pt\b**-\b**\fi%
\k*\a**\a**=\the\k*\dd*\k*\b**\b**=\the\k*\dd*}
\def\wavearcto#1(#2,#3)[#4]{\rlap{\toks0={#2}\toks1={#3}\calcnum*#1(#2,#3)%
\c*=\the\N*5sp\a*=.4\c*\a*=#4\a*\ifdim\a*<0pt\a*-\a*\fi%
\advance\c*\a*\N*\c*\divide\N*20\multiply\N*2\advance\N*-1\multiply\N*4%
\d**=#1\Lengthunit\dd*=.012\d**\ifdim\d**<0pt\d**-\d**\fi\L*=.25\d**%
\divide\d**\N*\divide\dd*\N*\L*=#4\L*\divide\L*\*ths%
\m*\N*\divide\m*2\dm*=\the\m*0sp\l*\dm*%
\sm*\n*\*one\loop\calcnormal*{#4}\calcparab*%
\*shl*{1}\advance\n*\*one\calcparab*%
\*shl*{1.3}\advance\n*\*one\calcparab*%
\*shl*{1}\advance\n*2%
\dd*-\dd*\ifnum\n*<\N*\repeat\n*\N*\shl**{0pt}}}
\def\triangarcto#1(#2,#3)[#4]{\rlap{\toks0={#2}\toks1={#3}\calcnum*#1(#2,#3)%
\c*=\the\N*5sp\a*=.4\c*\a*=#4\a*\ifdim\a*<0pt\a*-\a*\fi%
\advance\c*\a*\N*\c*\divide\N*20\multiply\N*2\advance\N*-1\multiply\N*2%
\d**=#1\Lengthunit\dd*=.012\d**\ifdim\d**<0pt\d**-\d**\fi\L*=.25\d**%
\divide\d**\N*\divide\dd*\N*\L*=#4\L*\divide\L*\*ths%
\m*\N*\divide\m*2\dm*=\the\m*0sp\l*\dm*%
\sm*\n*\*one\loop\calcnormal*{#4}\calcparab*%
\*shl*{1}\advance\n*2%
\dd*-\dd*\ifnum\n*<\N*\repeat\n*\N*\shl**{0pt}}}
\def\hr*#1{\clap{\xL*=\xscale\Lengthunit\vrule width#1\xL* height.1pt}}
\def\shade#1[#2]{\rlap{\Lengthunit=#1\Lengthunit%
\smov(0,#2.05){\hr*{.994}}\smov(0,#2.1){\hr*{.980}}%
\smov(0,#2.15){\hr*{.953}}\smov(0,#2.2){\hr*{.916}}%
\smov(0,#2.25){\hr*{.867}}\smov(0,#2.3){\hr*{.798}}%
\smov(0,#2.35){\hr*{.715}}\smov(0,#2.4){\hr*{.603}}%
\smov(0,#2.45){\hr*{.435}}}}
\def\dshade#1[#2]{\rlap{%
\Lengthunit=#1\Lengthunit\if#2-\def\t*{+}\else\def\t*{-}\fi%
\smov(0,\t*.025){%
\smov(0,#2.05){\hr*{.995}}\smov(0,#2.1){\hr*{.988}}%
\smov(0,#2.15){\hr*{.969}}\smov(0,#2.2){\hr*{.937}}%
\smov(0,#2.25){\hr*{.893}}\smov(0,#2.3){\hr*{.836}}%
\smov(0,#2.35){\hr*{.760}}\smov(0,#2.4){\hr*{.662}}%
\smov(0,#2.45){\hr*{.531}}\smov(0,#2.5){\hr*{.320}}}}}
\def\vdot{\rlap{\kern-1.9pt\lower1.8pt\hbox{$\scriptstyle\bullet$}}}
\def\vtimes{\rlap{\kern-3pt\lower1.8pt\hbox{$\scriptstyle\times$}}}
\def\vDot{\rlap{\kern-2.3pt\lower2.7pt\hbox{$\bullet$}}}
\def\vTimes{\rlap{\kern-3.6pt\lower2.4pt\hbox{$\times$}}}
\catcode`\*=12
\newcount\CatcodeOfAtSign
\CatcodeOfAtSign=\the\catcode`\@
\catcode`\@=11
\newcount\n@ast
\def\n@ast@#1{\n@ast0\relax\get@ast@#1\end}
\def\get@ast@#1{\ifx#1\end\let\next\relax\else%
\ifx#1*\advance\n@ast1\fi\let\next\get@ast@\fi\next}
\newif\if@up \newif\if@dwn
\def\up@down@#1{\@upfalse\@dwnfalse%
\if#1u\@uptrue\fi\if#1U\@uptrue\fi\if#1+\@uptrue\fi%
\if#1d\@dwntrue\fi\if#1D\@dwntrue\fi\if#1-\@dwntrue\fi}
\def\halfcirc#1(#2)[#3]{{\Lengthunit=#2\Lengthunit\up@down@{#3}%
\if@up\smov(0,.5){\arc[-][-]\arc[+][-]}\fi%
\if@dwn\smov(0,-.5){\arc[-][+]\arc[+][+]}\fi%
\def\lft{\smov(0,.5){\arc[-][-]}\smov(0,-.5){\arc[-][+]}}%
\def\rght{\smov(0,.5){\arc[+][-]}\smov(0,-.5){\arc[+][+]}}%
\if#3l\lft\fi\if#3L\lft\fi\if#3r\rght\fi\if#3R\rght\fi%
\n@ast@{#1}%
\ifnum\n@ast>0\if@up\shade[+]\fi\if@dwn\shade[-]\fi\fi%
\ifnum\n@ast>1\if@up\dshade[+]\fi\if@dwn\dshade[-]\fi\fi}}
\def\halfdashcirc(#1)[#2]{{\Lengthunit=#1\Lengthunit\up@down@{#2}%
\if@up\smov(0,.5){\dasharc[-][-]\dasharc[+][-]}\fi%
\if@dwn\smov(0,-.5){\dasharc[-][+]\dasharc[+][+]}\fi%
\def\lft{\smov(0,.5){\dasharc[-][-]}\smov(0,-.5){\dasharc[-][+]}}%
\def\rght{\smov(0,.5){\dasharc[+][-]}\smov(0,-.5){\dasharc[+][+]}}%
\if#2l\lft\fi\if#2L\lft\fi\if#2r\rght\fi\if#2R\rght\fi}}
\def\halfwavecirc(#1)[#2]{{\Lengthunit=#1\Lengthunit\up@down@{#2}%
\if@up\smov(0,.5){\wavearc[-][-]\wavearc[+][-]}\fi%
\if@dwn\smov(0,-.5){\wavearc[-][+]\wavearc[+][+]}\fi%
\def\lft{\smov(0,.5){\wavearc[-][-]}\smov(0,-.5){\wavearc[-][+]}}%
\def\rght{\smov(0,.5){\wavearc[+][-]}\smov(0,-.5){\wavearc[+][+]}}%
\if#2l\lft\fi\if#2L\lft\fi\if#2r\rght\fi\if#2R\rght\fi}}
\def\Circle#1(#2){\halfcirc#1(#2)[u]\halfcirc#1(#2)[d]\n@ast@{#1}%
\ifnum\n@ast>0\clap{%
\dimen0=\xscale\Lengthunit\vrule width#2\dimen0 height.1pt}\fi}
\def\wavecirc(#1){\halfwavecirc(#1)[u]\halfwavecirc(#1)[d]}
\def\dashcirc(#1){\halfdashcirc(#1)[u]\halfdashcirc(#1)[d]}
%
\def\xscale{1}
\def\yscale{1}
\def\Ellipse#1(#2)[#3,#4]{\def\xscale{#3}\def\yscale{#4}%
\Circle#1(#2)\def\xscale{1}\def\yscale{1}}
\def\dashEllipse(#1)[#2,#3]{\def\xscale{#2}\def\yscale{#3}%
\dashcirc(#1)\def\xscale{1}\def\yscale{1}}
\def\waveEllipse(#1)[#2,#3]{\def\xscale{#2}\def\yscale{#3}%
\wavecirc(#1)\def\xscale{1}\def\yscale{1}}
\def\halfEllipse#1(#2)[#3][#4,#5]{\def\xscale{#4}\def\yscale{#5}%
\halfcirc#1(#2)[#3]\def\xscale{1}\def\yscale{1}}
\def\halfdashEllipse(#1)[#2][#3,#4]{\def\xscale{#3}\def\yscale{#4}%
\halfdashcirc(#1)[#2]\def\xscale{1}\def\yscale{1}}
\def\halfwaveEllipse(#1)[#2][#3,#4]{\def\xscale{#3}\def\yscale{#4}%
\halfwavecirc(#1)[#2]\def\xscale{1}\def\yscale{1}}
\catcode`\@=\the\CatcodeOfAtSign

\title{The ambiguity-free four-dimensional Lorentz-breaking Chern-Simons action}
\author{F.A. Brito,$^{a}$ J.R. Nascimento,$^{b,c}$ E. Passos$^{b}$, and A. Yu. Petrov,$^{b}$}
\affiliation{\small{$^a$Departamento de F\'\i sica, Universidade
Federal de Campina Grande, Caixa Postal 10071,
58109-970 Campina Grande, Para\'\i ba, Brazil\\
$^b$Departamento de F\'\i sica, Universidade Federal da Para\'\i ba,
Caixa Postal 5008, 58051-970 Jo\~ao Pessoa, Para\'\i ba,
Brazil\\
$^c$Instituto de F\'\i sica, Universidade de S\~ao Paulo, Caixa
Postal 66318, 05315-970 S\~ao Paulo, SP, Brazil}}
\date{\today}
\begin{abstract}
The four dimensional Lorentz-breaking finite and determined
Chern-Simons like action is generated as a one loop perturbative
correction via an appropriate Lorentz-breaking coupling of the gauge
field with the spinor field. Unlike the known schemes of
calculations, within this scheme this term is found to be
regularization independent.
\end{abstract}
 \maketitle


The possibility of the Lorentz and CPT symmetries being violated in
the nature has been considered in recent years
\cite{0,1,2,3,4,5,6,7,biref,8,9,10,11,11.2,12,13,14}. Several
theoretical investigations have pointed out that these symmetries
can be approximate. In such studies one has mainly considered that
the realization of this violation can be obtained in QED by adding
the Chern-Simons-like term ${\cal
L}_{CS}=\frac{1}{2}k_\mu\epsilon^{\mu\alpha\beta\gamma}F_{\alpha\beta}A_\gamma$,
with $k_{\mu}$ being a constant quadrivector characterizing the
preferred direction of the space-time, to the photon sector, and
another term which is a CPT-odd term for fermion sector, i.e.,
$\bar{\psi}{\bs}\gamma_5\psi$ with a constant quadrivector $b_\mu$
introducing CPT symmetry breaking. As one knows, this extension of
the QED does not break the gauge symmetry of the action and
equations of motion but it does modify the dispersion relations for
different polarization of photons and Dirac's spinors. The
Chern-Simons-like term is known to have some important implications,
such as birefringence of light in the vacuum \cite{biref}. Many
interesting investigations in the context of Lorentz-CPT violation
have appeared recently in the literature. For instance, several
issues were addressed, such as \v{C}erenkov-type mechanism called
``vacuum \v{C}erenkov radiation'' to test the Lorentz symmetry
\cite{ptl}, changing of gravitational redshifts for differently
polarized Maxwell-Chern-Simons photons \cite{kkl}, evidence for the
Lorentz-CPT violation from the measurement of CMB polarization
\cite{bjx}, supersymmetric extensions \cite{bch}, breaking of the
Lorentz group down to the little group associated with $k_\mu$
\cite{hle} and magnetic monopoles inducing electric current
\cite{brz}. Among these developments one of the most interesting and
controversial result is the dynamical origin of the parameters
$k_\mu$ and $b_\mu$ present in the Lorentz and CPT symmetry breaking
obtained when we integrate over the fermion fields in the modified
Dirac action. The result is the induction of the Chern-Simons term
via radiative corrections which may lead to a relation between
$k_\mu$ and $b_\mu$.

The induction of the Chern-Simons-like Lorentz-CPT violating term,
${\cal L}_{CS}$, is one of the most important result in the study of
the Lorentz symmetry violation \cite{2,3,7}. This term, naturally
emerges as a perturbative correction in the theory suggested in
\cite{3} as a possible extension of QED by an axial-vector term
\ben\label{QEDext} {\cal L}=\bar{\psi}( i \pls - m )\psi -
\bar{\psi} \bs \gamma_{5}\psi - e\bar{\psi} \as \psi. \een Carrying
out the integration over fermions, the relation between the
coefficients $k_{\mu}$ and $b_{\mu}$ could be obtained in terms of
some loop integrals being some of them divergent. Therefore one has
to implement some regularization to calculate these integrals, thus,
the constant $C$ relating the coefficients as $k_{\mu}=C b_{\mu}$
turns out to be dependent on the regularization scheme used
\cite{bpp}. The ambiguity of the results manifested in the
dependence on the regularization scheme has been intensively
discussed in the literature. Several studies have shown that $C$ can
be found to be finite but undetermined \cite{23, 24, 25, 26, 27}. {
Astrophysical observations impose very stringent experimental bounds
on $k_{\mu}$ indicating it vanishes. Since the coefficient $k_{\mu}$
of the radiatively induced Chern-Simons term depends on $b_\mu$ it
is natural to expect that the constant $b_\mu$ can also suffer an
experimental bound in this framework. However, if ambiguities are
present there is no way to know for sure about what extent the
experimental bounds constraint the constant $b_\mu$, because $C$ is
simply undetermined. In other words, we cannot define the fate of
the constant that is responsible for the Lorentz and CPT violation
in the fermion sector by simply measuring $k_\mu$. In the following
we are going to extend the well studied Lagrangian (\ref{QEDext}) in
attempting to shed some light on the issue of inducing Chern-Simons
term with no ambiguities.}

In this letter we propose an extension of the usual theory through
new chiral couplings that can eliminate such ambiguities. So, let us
introduce an extension of the usual Lagrangian (\ref{QEDext}) as in
the following \ben {\cal L}=\bar{\psi}( i \pls - m )\psi -
\bar{\psi} \bs (1 + \gamma_{5}) \psi - e\bar{\psi} \as (1 -
\gamma_{5}) \psi. \label{QEDQ}\een {In order to show that the above
Lagrangian  preserves gauge invariance we rewrite it in terms of new
gauge fields, a vector field and an axial field, in the form \cite{REI}\ben
{\cal L}= \bar{\psi}( i \pls - m  -  \cvs -\cas\gamma_{5})
\psi,\label{QEDQQ}\een where the vector and axial gauge fields are
defined as $\cvs=\bs+e \as$ and $\cas=\bs-e \as$, respectively. In
this sense one can understand $b_\mu$ and $A_\mu$ as  $R$- and
$L$-handed external fields. Hence, this Lagrangian is invariant
under the local vector gauge transformation $U_{V}(1)$
\ben &&\psi\rightarrow \exp[i{\alpha}(x)]\psi,\nonumber \\
&&\bar\psi\rightarrow \bar\psi\exp[-i{\alpha}(x)],\\
&&{\cal V}_{\mu}\rightarrow {\cal
V}_{\mu}-\partial_{\mu}\alpha(x)\nonumber. \een For massless
fermions ($m=0$) the Lagrangian (\ref{QEDQQ}) also admits the local
axial gauge transformation $U_{A}(1)$
\ben &&\psi\rightarrow \exp[i{\gamma_5\beta}(x)]\psi,\nonumber \\
&&\bar\psi\rightarrow \bar\psi\exp[i{\gamma_5\beta}(x)],\\
&&{\cal A}_{\mu}\rightarrow {\cal
A}_{\mu}-\partial_{\mu}\beta(x)\nonumber, \een so that, altogether
the Lagrangian (\ref{QEDQQ}) has the combined gauge symmetry
$U_{V}(1)\times U_{A}(1)$. Therefore, we can conclude that the
vector and axial currents are conserved separately, i.e.,
$\partial^{\mu}j_{\mu}=0$ and $\partial^{\mu}j_{\mu}^{5}=0$ (for
$m=0$).}

The couplings we propose in the theory yields a special gauge
invariant and Lorentz-CPT violating theory that we call `extended
chiral QED', where divergences among loop integrals are canceled.
The extension is based on the observation that one can extend and
impose some restrictions on the gauge invariant and CPT-Lorentz
violating Lagrangian (\ref{QEDext}) by replacing $\bs$ and $\as$
according to the transformations \ben\bs\gamma_5\to \bs (1 +
\gamma_{5}),
\\ \as\to \as (1 - \gamma_{5}).\een One can also show that other combinations just
changing signals in the interacting terms above produces either only
divergent integrals or mixture of divergent and finite integrals.
Note that both $b_\sigma$ and $A_\sigma$ act as external fields with
opposite chirality interacting with fermion fields. As we shall show
below, in this model the divergences are canceled and no
regularization scheme is required. As we just anticipated, by
combining other signals of $\gamma_5$ in (\ref{QEDQ}), one could
also have interactions with same chirality, but in this case the
divergences would persist.

The one loop effective action  $S_{eff}[b,A]$ of the gauge field
$A_{\mu}$ in this theory can be expressed in the form of the
following functional trace \be S_{eff}[b,A]=-i\,{\rm Tr}\,\ln(\ps- m
- \bs(1+\gamma_5)-e \as (1-\gamma_5) ). \ee This functional trace
can be represented as
$S_{eff}[b,A]=S_{eff}[b]+S_{eff}^{\,\prime}[b,A]$, where the first
term is $S_{eff}[b]=-i\,{\rm Tr}\ln(\ps- m - \bs(1+\gamma_5))$,
which does not depend on the gauge field. The only nontrivial
dynamics is governed by the second term $S_{eff}^{\,\prime}[b,A]$,
which is given by the following power series \be\label{ea}
S_{eff}^{\,\prime}[b,A]=i\,{\rm Tr} \sum_{n=1}^{\infty}\frac1n
\Biggl[\frac1{\ps- m - \bs(1+\gamma_5)}\,e \as(1-\gamma_5)\Biggr]^n.
\ee To obtain the Chern-Simons term we should expand this expression
up to the second order in the gauge field \be \label{Ef1}
S_{eff}^{\,\prime}[b,A]=S_{eff}^{(2)}[b,A]+\ldots. \ee The dots in
(\ref{Ef1}) stand for the terms of higher orders in the gauge field.
Here \be \label{efa2} S_{eff}^{(2)}[b,A]=-\frac{ie^{2}}{2}{\rm
Tr}\bigl[S_b(p)\;\as(1-\gamma_5)\;S_b(p)\;\as(1-\gamma_5)\bigl], \ee
where $S_b(p)$ is the $b^{\mu}$-dependent propagator of the theory
defined as \bea S_b(p)=\frac{i}{ \ps- m - \bs(1+\gamma_5)}, \eea
which can be expanded in the following series in $b^{\mu}$ \bea
\label{expa} S_b(p)=S(p)+S(p)(-i\bs(1+\gamma_5))S(p)+\ldots, \eea
with $S(p)$ being the usual fermion propagator. Thus, using the
derivative expansion method \cite{de} (see also \cite{bpp}), the one
loop effective action (\ref{efa2}) can be equivalently rewritten as
\bea \label{efa2a} S_{eff}^{(2)}[b,A]=\int d^4x
\Pi^{\lambda\mu\nu}A_{\mu}\pa_{\nu}A_{\lambda}, \eea with the one
loop self-energy given by \bea\label{efa2aa}
&&\Pi^{\lambda\mu\nu}=-\frac{ie^2}{2}\int
\frac{d^4p}{(2\pi)^4}\times\nonumber\\&&\times\;{\rm
tr}\left[S_b(p)\gamma^{\mu}(1-\gamma_5)S_b(p)\gamma^{\lambda}S_b(p)\gamma^{\nu}(1-\gamma_5)
\right].\eea Substituting into (\ref{efa2aa}) the expansion of the
propagator (\ref{expa}), taking into account the fact that
$(\gamma_5)^2=1$ and applying the following trace \bea {\rm
tr}(\gamma^{\lambda}\gamma^{\mu}\gamma^{\nu}\gamma^{\rho}\gamma_5)=4i\epsilon^{\lambda\mu\nu\rho},
\eea we can write down the simple expression for
$\Pi^{\lambda\mu\nu}$ \bea \label{proj}
\Pi^{\mu\nu\lambda}=8ie^2m^2\,b_{\rho}\int\frac{d^4p}{(2\pi)^4}\frac{N^{\mu\nu\lambda}}{(p^2-m^2)^4},
\eea where \bea
N^{\mu\nu\lambda}=2\,\epsilon^{\mu\nu\rho\theta}p^{\lambda}p_{\theta}
+ \epsilon^{\mu\nu\rho\lambda}(p^{2}-m^{2}).\eea  The fundamental
fact is that, unlike the results obtained earlier
\cite{bpp,7,AA,N3,gnpp}, this result is manifestly finite and does
not require any regularization. The exact, regularization
independent, value for $\Pi^{\lambda\mu\nu}$ is \bea
\Pi^{\lambda\mu\nu}=\epsilon^{\lambda\mu\nu\rho}b_{\rho}\frac{e^2}{3\pi^2}.
\eea Thus, the effective action (\ref{efa2a}) acquires the familiar
form \bea S=\frac{1}{2}\int
d^4x\epsilon^{\lambda\mu\nu\rho}k_{\rho}A_{\mu}F_{\nu\lambda},\eea
with the following relation between the constant quadrivectors
$k_\rho$ and  $b_\rho$ \bea
\label{cs_k}k_{\rho}=\frac{e^2}{3\pi^2}b_{\rho}.\eea

This relation is finite and determined.  Thus confirming our
attempting, our Lorentz-CPT violating `extended quiral QED' theory,
radiatively induces a Chern-Simons term which displays uniqueness of
the result due to absence of the divergences achieved without
imposing any regularization. A natural justification of our result
can be attributed to the fact that the considered model in
(\ref{QEDQ}) represents a more complete description of what the
usual model (\ref{QEDext}). We see that in the derivative expansion method, the model (\ref{QEDext}) gets 
only three contributions, whereas in our model (\ref{QEDQ}) twelve contributions are present.  We attribute this to the symmetry in  the new quiral coupling terms in the
Lagrangian (\ref{QEDQ}) that introduces new vertices and new insertions
into the fermion propagator that can add new one loop contributions
in relation to the usual model (\ref{QEDext}), which are responsible
for divergence cancellations.  Now let us consider the Feynman
rules in the following. The fermion propagator is

\vspace*{1mm}

\hspace{2cm} \Lengthunit=1.2cm
\GRAPH(hsize=3){\Linewidth{.6pt}\lin(1,0)\ind(15,0){\;\;\;\;\;\;\;\;\;\;\;\;=\frac{i(\ps+m)}{p^{2}-m^{2}}}
}

\vspace*{1mm}

\noindent and the coefficients for CPT violation lead to insertions
into the fermion propagator

\vspace*{1mm}

\hspace{2cm}
\Lengthunit=1.2cm
\GRAPH(hsize=3){\ind(5,0){\bullet}\ind(5,0){\times}\lin(1,0)\ind(18,0){\;\;\;=-i\bs\gamma_5}
}

\vspace*{1mm}

\hspace{2cm}
\Lengthunit=1.2cm
\GRAPH(hsize=3){\ind(5,0){\times}\lin(1,0)\ind(18,0){=-i\bs}
}

\vspace*{1mm}

The chiral and usual fermion-photon vertex are respectively

\vspace*{1mm}

\hspace{2cm}
\Lengthunit=1.2cm
\GRAPH(hsize=3){\ind(5,0){\bullet}\lin(1,0)\mov(.5,0){\wavelin(0,.7)}\ind(22,0){=ie\gamma^{\mu}\gamma_5}
}

\vspace*{1mm}

\hspace{2cm}
\Lengthunit=1.2cm
\GRAPH(hsize=3){\lin(1,0)\mov(.5,0){\wavelin(0,.7)}\ind(18,0){\;\;\;\;\;\;\;\;\;=-ie\gamma^{\mu},}
}

\noindent where $e$ is the fermion charge and $\mu$ the space-time
index on the photon line. \vspace*{1mm}
 We find that the following diagrams contribute to the
two-point function:

\vspace*{3mm}

\hspace{.5cm}
\Lengthunit=1.2cm
\GRAPH(hsize=2){\wavelin(.5,0)\mov(1,0){\Circle(1)}\mov(1.5,0){\wavelin(.5,0)}\ind(10,-5){\bullet}\ind(10,-5){\times}
\ind(10,-10){(a)}
}
\hspace{.5cm}
\Lengthunit=1.2cm
\GRAPH(hsize=2){\wavelin(.5,0)\mov(1,0){\Circle(1)}\mov(1.5,0){\wavelin(.5,0)}\ind(10,5){\bullet}\ind(10,5){\times}
\ind(10,-10){(b)}
}

\vspace*{3mm}

\hspace{.5cm} \Lengthunit=1.2cm
\GRAPH(hsize=2){\wavelin(.5,0)\mov(1,0){\Circle(1)}\mov(1.5,0){\wavelin(.5,0)}\ind(10,-5){\times}\ind(15,0){\bullet}
\ind(10,-10){(c)} } \hspace{.5cm} \Lengthunit=1.2cm
\GRAPH(hsize=2){\wavelin(.5,0)\mov(1,0){\Circle(1)}\mov(1.5,0){\wavelin(.5,0)}\ind(10,5){\times}\ind(15,0){\bullet}
\ind(10,-10){(d)} }

\vspace*{3mm}

\hspace{.5cm}
\Lengthunit=1.2cm
\GRAPH(hsize=2){\wavelin(.5,0)\mov(1,0){\Circle(1)}\mov(1.5,0){\wavelin(.5,0)}\ind(10,-5){\times}\ind(5,0){\bullet}
\ind(10,-10){(e)}
}
\hspace{.5cm}
\Lengthunit=1.2cm
\GRAPH(hsize=2){\wavelin(.5,0)\mov(1,0){\Circle(1)}\mov(1.5,0){\wavelin(.5,0)}\ind(10,5){\times}\ind(5,0){\bullet}
\ind(10,-10){(f)}
}

\vspace{2mm}

\hspace{.5cm}
\Lengthunit=1.2cm
\GRAPH(hsize=2){\wavelin(.5,0)\mov(1,0){\Circle(1)}\mov(1.5,0){\wavelin(.5,0)}\ind(10,-5){\bullet}\ind(10,-5){\times}
\ind(4,0){\bullet}\ind(14,0){\bullet}
\ind(10,-10){(g)}
}
\hspace{.5cm}
\Lengthunit=1.2cm
\GRAPH(hsize=2){\wavelin(.5,0)\mov(1,0){\Circle(1)}\mov(1.5,0){\wavelin(.5,0)}\ind(10,5){\bullet}\ind(10,5){\times}
\ind(4,0){\bullet}\ind(14,0){\bullet}
\ind(10,-10){(h)}
}

\vspace*{1mm}

The one loop self-energy tensor for the graphics in the first and
the second columns above are respectively \bea\label{t2}
&&\!\!\!\!\!\!\!\!\!\Pi^{\mu\nu}_{a,c,e,g}(q)\!=\!e^{2}\!\!\int\!\!\frac{d^{4}p}{(2\pi)^{4}}{\rm
tr}\bigl[\gamma^{\mu}S(p)\gamma^{\nu}S(p\!+\!q)\bs\gamma_{5}S(p\!+\!q)-\nonumber\\
&&\!\!\!\!\!\!\!\!\!\gamma^{\mu}S(p)\gamma^{\nu}\gamma_{5}S(p\!+\!q)\bs
S(p\!+\!q)\!-\!\gamma^{\mu}\gamma_{5}S(p)\gamma^{\nu}S(p\!+\!q)\bs S(p\!+\!q)\nonumber\\
&+&\gamma^{\mu}\gamma_{5}S(p)\gamma^{\nu}\gamma_{5}S(p\!+\!q)\bs\gamma_{5}S(p\!+\!q)\bigl]\nonumber
\eea and \bea\label{t1}
&&\!\!\!\!\!\!\!\!\!\Pi^{\mu\nu}_{b,d,f,h}(q)\!=\!e^{2}\!\!\int\!\!\frac{d^{4}p}{(2\pi)^{4}}{\rm
tr}\bigl[\gamma^{\mu}S(p)\bs\gamma_{5}S(p)\gamma^{\nu}S(p\!+\!q)-\nonumber\\
&&\!\!\!\!\!\!\!\!\!\gamma^{\mu}S(p)\bs
S(p)\gamma^{\nu}\gamma_{5}S(p\!+\!q)\!
-\!\gamma^{\mu}\gamma_{5}S(p)\bs S(p)\gamma^{\nu}S(p\!+\!q)\nonumber\\
&+&\gamma^{\mu}\gamma_{5}S(p)\bs\gamma_{5}S(p)\gamma^{\nu}\gamma_{5}S(p\!+\!q)\bigl],\nonumber
\eea where $S(p)$ is the usual fermion propagator. Considering
the self-energy tensor $\Pi^{\mu\nu}_{a,c,e,g}(q)$ or
$\Pi^{\mu\nu}_{b,d,f,h}(q)$, we can find finite and
infinite contributions that should be regularized. Thus integration
momentum in each self-energy tensor produces finite but undetermined
results, although  the total integration momentum, under appropriate regularization, leads to a result that coincides
with the finite and regularization independent (determined) Chern-Simons term coefficient
(\ref{cs_k}) obtained within the derivative expansion method. Thus, in our extension we have an improved theory
whose quantum calculations give us a finite non-ambiguous answer at
one loop level, as long as we apply the derivative expansion method.

We summarize our results as follows. In our present analysis, we
have shown that due to the presence of chiral couplings
preserving a special gauge invariance in the Lagrangian density, by using the derivative 
expansion method, we
can avoid ambiguities. Transformations under this special symmetry
maintain the gauge invariance that occurs both in the Lagrangian
density and the action, which yields a new result finite and
determined for the Chern-Simons coefficient. Differently of the
usual theory studied in the literature, where the breaking of the
gauge invariance occurs in the quantum level, in the present case
the gauge invariance is maintained at both {\it classical and
quantum level} because of the absence of regulators, and the problem of
divergences and ambiguities does not exist. The radiatively induced
Chern-Simons term is not gauge invariant but its spacetime integral
in the action is. Equivalently, it means that the Chern-Simons term
is invariant only in zero momentum \cite{23}. The regularization
schemes such as Pauli-Villars and gauge invariant dimensional
regularization, normally used to regularize divergent integrals in
the usual theory, requires gauge invariance at all momenta which
exclude a priori the possibility of inducing a Chern-Simons term.
Since Chern-Simons terms are gauge invariant only in zero momentum,
thus in order to have a finite and determined Chern-Simons term
coefficient $k_\mu$, one would have to use regularization schemes
that could involve gauge invariance only at zero momentum too. A
natural development in order to solve this problem would be to find
the possibility of computing the Chern-Simons coefficient by not
using any regularization scheme. Thus one possibility that we have
chosen in this investigation was to look for a theory where the
divergent one loop integral graphs at momentum space could be
eliminated such that the remaining integrals of the theory were
finite. We were able to find a theory that really does the job. The
theory we found invokes new quiral couplings that preserve a special
gauge invariance in addition to violate Lorentz and CPT symmetries.
It comprises a theory where both $b_\sigma$, present in the
axial-vector term, and the gauge vector potential $A_\sigma$ act as
external fields with opposite chirality interacting with fermion
fields that renders a complete cancelation of divergences and in
turn no regularization scheme is required.

{\bf Acknowledgements.} This work was partially supported by Conselho Nacional
de Desenvolvimento Cient\'{\i}fico e Tecnol\'{o}gico (CNPq). The
work by A. Yu. P. has been supported by CNPq-FAPESQ DCR program,
CNPq project No. 350400/2005-9.

\end{document}